\documentclass[12pt,preprint]{aastex}
\usepackage{epsfig}

\begin{document}

\title{$^{12}$C/$^{13}$C ratio in planetary nebulae from the IUE archives}
\author{R.H. Rubin\altaffilmark{1,2}}
\author {G.J. Ferland\altaffilmark{3}}
\author{E.E. Chollet\altaffilmark{1}}
\author{R. Horstmeyer\altaffilmark{1}}

\email{rubin@cygnus.arc.nasa.gov; gary@pa.uky.edu}

\altaffiltext{1} {NASA/Ames Research Center, Moffett Field, CA
94035-1000.}
\altaffiltext{2} {Orion Enterprises, M.S. 245-6, Moffett Field, CA
94035-1000}
\altaffiltext{3} {University of Kentucky, Department of Physics and
Astronomy, Lexington, KY 40506.}

\begin{abstract}
  We investigated the abundance ratio of $^{12}$C/$^{13}$C in planetary
nebulae by examining 
emission lines arising from 
\ion{C}{3} $2s2p~^3P_{2,1,0} \rightarrow 2s^2~^1S_0$.       
Spectra were retrieved from the International Ultraviolet
Explorer archives, and multiple spectra of the same object were coadded to
achieve improved signal-to-noise.  
The $^{13}$C hyperfine structure line 
at 1909.6~\AA\ was detected in NGC~2440.
The $^{12}$C/$^{13}$C ratio was found to be $\sim$4.4$\pm$1.2. 
In all other objects, we provide an upper limit for the flux of the
1910~\AA\ line.
For 23 of these sources,
a lower limit for the $^{12}$C/$^{13}$C
ratio was established.  
The impact on our current
understanding of stellar evolution is discussed.

	The resulting high signal-to-noise \ion{C}{3} spectrum
helps constrain the atomic physics of the line formation process.
Some objects have the measured 1907/1909 flux ratio outside the 
low-electron density theoretical limit for $^{12}$C.
A mixture of $^{13}$C with $^{12}$C helps to close the
gap somewhat.
Nevertheless, some observed 
1907/1909 flux ratios still appear too high to 
conform to the presently predicted limits.
It is shown that this limit,
as well as the 1910/1909 flux ratio, 
are predominantly influenced
by using the standard partitioning among the collision strengths
for the multiplet $^1S_0$--$^3P_J$ 
according to the statistical weights.
A detailed calculation for the fine structure collision
strengths between these individual levels would be valuable.

\end {abstract}


\section{Introduction}

  The relative abundances of $^{12}$C and $^{13}$C in evolved stars
present a strong challenge to our current understanding of stellar
nucleosynthesis.  A growing body of observational evidence suggests that
standard stellar models greatly overestimate the ratio of $^{12}$C to
$^{13}$C on the red giant branch (RGB) and later evolution. 
In a careful study of
the literature, \citet{cha98} have shown that a full 96\% of evolved stars
exhibit a $^{12}$C/$^{13}$C ratio below 15, considerably lower than models
predict ($^{12}$C/$^{13}$C between 20~-- 30).  
On the other hand,
cool bottom processing (CBP), which employs a deep mixing
mechanism below the standard convective envelope during the red giant
phase \citep{sac99}, predicts $^{12}$C/$^{13}$C ratios $\leq$20 for
low-mass post-RGB stars \citep{boo99}.  Using millimeter wave observations
of $^{12}$CO and $^{13}$CO, \citet{pal00} found several examples of
planetary nebulae (PNs) with small $^{12}$C/$^{13}$C ratios, as low as 9
for NGC 7293. 
Balser et~al.\ (2002) in a similar study measured
$^{12}$CO and $^{13}$CO in molecular clouds associated with PNs.
They found the range of $^{12}$C/$^{13}$C from 2.2--31 in 9 PNs.
The extreme low value 2.2$\pm$0.03 was found for the PN M1--16.
Developing another approach to obtaining these abundance ratios
is very valuable to corroborate these low values.

  \citet{cle85} 
pointed out that the $^{12}$C/$^{13}$C
ratio could be determined using an extremely weak \ion{C}{3}
$^{13}$C line at
1909.597~\AA. 
This line is a hyperfine induced transition that arises from
$2s2p~^3P_0 \rightarrow 2s^2~^1S_0$,       
which is strictly forbidden in $^{12}$C. 
By comparing
the flux of this line to the nearby bright \ion{C}{3} lines 
$2s2p~^3P_{2,1} \rightarrow 2s^2~^1S_0$       
at 1906.683~\AA\ and 1908.734~\AA\ respectively, one can find the
$^{12}$C/$^{13}$C ratio. 
Using the Hubble Space Telescope (HST),
\citet{cle97} found a ratio of 15$\pm$3 in NGC 3918 and 21$\pm$11 in SMC
N2, as well as a very marginal $7^{+14}_{-3}$ for LMC N122. \citet{pal02},
also using HST, reported a non-detection in NGC 3242, showing
$^{12}$C/$^{13}$C~$\geq$ 38.

  When studying faint lines in extended sources, the light gathering power
of the instrument is extremely important.  
The large IUE aperture, an $\sim$20$''$$\times$$\sim$10$''$  $'$rounded 
rectangle$'$
(also described as an oval or race-track pattern), allowed a
high throughput for extended objects like PNs.  
The extensive archival
database and New Spectral Image Processing System (NEWSIPS) allow us to
effectively study the 
1909.597~\AA\ 
$^{13}$C line with IUE. 
Using these data, coadding multiple spectra when available to get 
a higher signal-to-noise ratio (S/N),  
we attempt to measure or provide upper limits for the flux of this line.
In conjunction with our measurements of fluxes in the adjacent ``main"
lines, we derive or set limits for the
$^{12}$C/$^{13}$C ratio in most of the sample of PNs.

        In section 2, we discuss the IUE database and our flux measurements.
The analysis tools to interpret the observations are introduced in
section 3, with the population rate equations.
In section 4, the flux ratios are analyzed in terms of the 
$^{12}$C/$^{13}$C ratio and the electron density ($N_e$).
Section~5 contains a discussion and conclusions.
        
\section{IUE Observations}

  The data were obtained from the IUE archives. 
For most sources, only short wavelength prime (SWP) camera spectra with the 
large aperture were used, because the long wavelength spectra taken with both
the prime camera (LWP) and redundant camera (LWR) are much noisier at this 
wavelength.  
The spectral resolution of high dispersion SWP, LWP, or LWR
spectra at these wavelengths is $\sim$0.2~\AA,
sufficient to separate the weak $^{13}$C line from the bright
line at 1908.7~\AA.

	We included 
all promising PNs observed with high dispersion in the large aperture.
Multiple observations of the same object were coadded point by point, 
weighted by the inverse of the total noise in the continuum,
to improve the S/N, 
using our own analysis program.
We note that the
inverse noise 
and exposure time are not linearly related in IUE spectra.  
Figure~1 shows a plot of inverse noise vs.\ integration time for
NGC~3918 SWP spectra.  
Though the
data quality, 
as gauged by inverse noise,
generally improves with increased time, the quality is not
a linear relation of either exposure time or the square root of the
exposure time. 
The IUE on-line ``Frequently Asked Questions" 
(http://archive.stsci.edu/cgi-bin/faq.cgi?mission=IUE\#40)
recommends inverse noise
weighting, and our data show that weighting by the inverse noise generates
better S/N than weighting by the time or by the square root of the time.

\begin{figure}[top]
\epsscale{.9}
\plotone{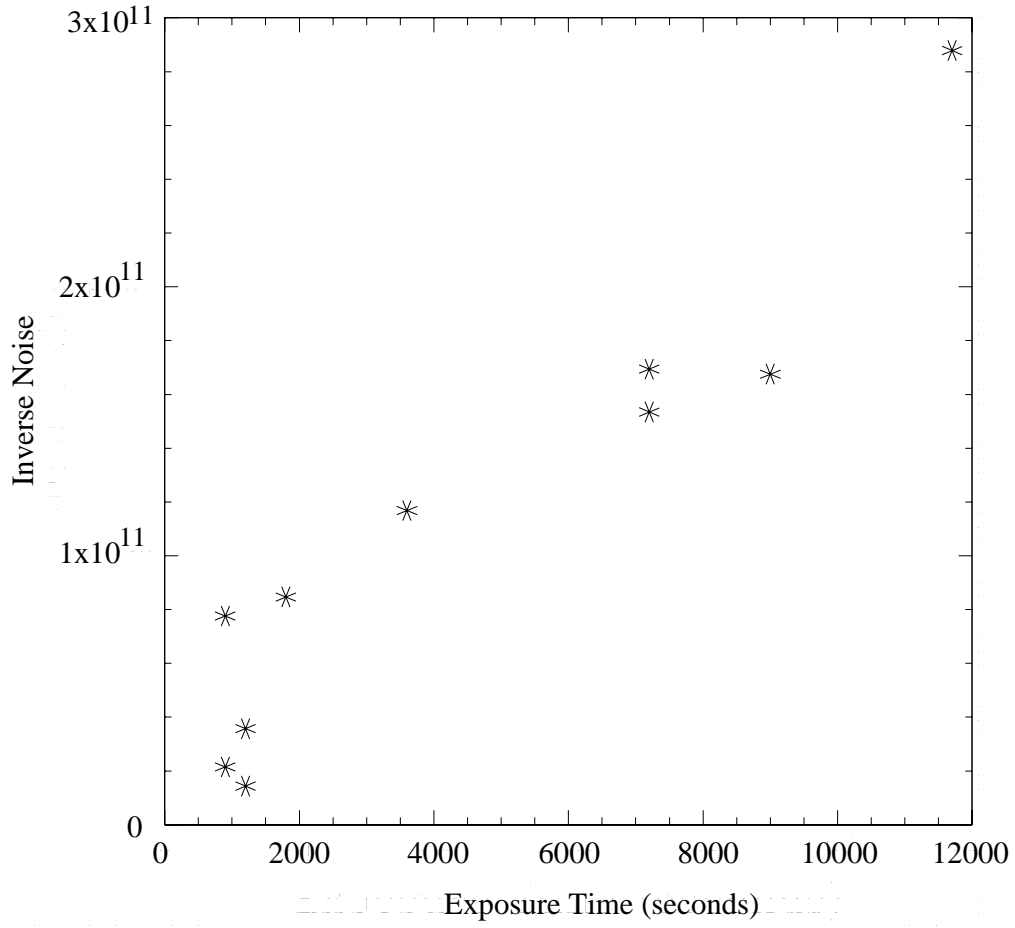}
\caption{The inverse noise 
(the reciprocal of the sum of noise points from 
$\sim$1910.5--1920~\AA)
versus the exposure time
for the various
SWP high dispersion, large aperture  
IUE spectra of NGC~3918.
Units in the ordinate are relative; 
a larger value indicates
less noisy data. \label{fig1}}
\end{figure}

        The ISAP\footnote{ISO Spectral Analysis Package (ISAP) 
is a joint development by the LWS and SWS Instrument Teams and Data
Centers.  
Contributing institutes are CESR, IAS, IPAC, MPE, RAL, \& SRON.}
routine Line Fit was used to fit a three-component 
Gaussian to the three lines of the \ion{C}{3} multiplet
in order to determine the line fluxes 
or to set an upper limit when a line was not detected.
The number of free parameters were limited by
constraining the full width half maximum (FWHM)
to be that of the 1907~\AA\ line.
Only one degree of freedom in the determination of wavelength
was allowed by requiring the fit to conform to 
the known $\Delta$$\lambda_{vac}$ between the transitions~--
that is an offset of 2.051 and 2.914~\AA\ for
the 1909 and 1910 lines relative to the 1906.683~\AA\ line.
We use a quadratic continuum fit to the line-free data.
Most of the IUE spectra that we found useful are SWP spectra.
For fitting the continuum baseline in these, we avoid the data 
points at lower wavelength than the 1907~\AA\ line where there
is an echelle interorder ``splice" (at $\sim$1905.4) and the 
fluctuations in the continuum
intensity  become much larger than those on the red side.

        We used both Gaussian-fitting 
and direct integration routines to measure the net
line flux above the continuum.  
Because the filled large aperture introduced a spectral
impurity into the spectrograph, the line profiles have flatter
tops and less extended bases (e.g., are more ``trapezoidal")
than the Gaussian fits.
For this reason and because some of the sources have
lines that depart in other ways from Gaussian,
we prefer to use the direct integration routines 
(the moment option in ISAP) to measure the line flux.
We also confirmed our line flux results using 
IRAF\footnote{IRAF is distributed by NOAO,
which is operated by AURA, under cooperative agreement with NSF.}
splot with the e-option for direct integration.

        The root-mean-square (rms) fluctuations in the continuum 
flux were determined from a region clear of lines and obvious 
spikes longward of 1910~\AA\ and the 
FWHM determined from the 1906.7~\AA\ line.
  From this, we determine the value 3$\sigma$, which is our threshold for
judging a flux measurement a detection or not.
This 3$\sigma$ value is also used to set upper limits on the flux for the
non-detections.
The results of our measurements are presented in Table~1.
Column~1 has the source name;
columns 2 and 3 the J2000 positions; columns 4, 6, and 8 have
the line flux measurements F(1906.7), F(1908.7), and F(1909.6) in
erg~cm$^{-2}$~s$^{-1}$.
Column~5 has the 1$\sigma$ uncertainty associated with each 
of the line fluxes while column 7 is 3 times this value in
order to judge whether or not the hyperfine line has been detected.
The FWHM from the Gaussian fit of the coadded, unsaturated 
spectra are in column~9.
The entries in the remaining columns are quantities derived 
from these data and will be discussed later.

	For many of the sources, there is a second line in Table~1
where more IUE spectra were coadded with those used to produce
the first-line results (with {\it only unsaturated data}).  
These added spectra are saturated in either the 1907 or 1909 lines.
Because of the limited dynamic range of IUE, long exposures,
particularly  with the SWP camera, of bright PNs will readily saturate 
(overexpose) these strong \ion{C}{3} lines.
However, there is not a  saturation 
issue with the 1910 line and adjacent
continuum.
Using these additional data, we are able to test if there are now any
statistically significant detections of the 1910 line by comparing
F(1909.6) with the 3$\sigma$ uncertainty, which is generally
smaller than the entry on line 1.
Although the resulting flux and FWHM in the main lines are not reliable, 
we are able to
use these lines as a fiducial wavelength indicator only to
fit the data in the vicinity of the 1910~\AA\ line with a single
Gaussian that has a FWHM assumed to be the same as on line 1.
As will be discussed (\S4.2), the only object with a significant detection
of the 1910 line is NGC~2440. 
Even with the additional data entered on line 2, we found no other 
significant detections of this line.

\begin{deluxetable}{ccccccccccccc}
\rotate
\tabletypesize{\scriptsize}
\tablecaption{\ion{C}{3} Line Fluxes and Derived Parameters \label{tbl-1}}
\tablewidth{0pt}
\tablehead{
\colhead{Object} & \colhead{~~RA ~~~J2000} & \colhead{DEC} & \colhead{F(1906.7)} &
\colhead{1$\sigma$ error} & \colhead{F(1908.7)} & \colhead{3$\sigma$}  
& \colhead{F(1909.6)} & \colhead{FWHM} & \colhead{1907/1909} &\colhead{$N_e$ cm$^{-3}$}
& \colhead{$r$ ($^{12}$C/$^{13}$C)}
}
\startdata
% Object          RA            DEC      1907(m,ns) 1 sigma(ns) 1909(m,ns) 3 sigma(ns) 1910(g,ns) FWHM(m) 1907/1909  Ne 12C/13C
% Object          RA            DEC                 1 sigma(s)             3 sigma(s)  1910(g,s)
PN G108.2-76.1 & 00 37 16.0   & -13 43 00  & 2.79E-12 & 3.42E-14 & 1.93E-12 & 1.03E-13 & -5.24E-14 & 0.3670 & 
1.45 &1580--2120 &$>$3 \\
& & &          & 2.04E-14 &          & 6.11E-14 & 3.00E-14 &        & & & \\

IC 351         & 03 47 32.9 & +35 02 48 & 1.71E-12 & 7.91E-14 & 1.40E-12 & 2.37E-13 & 3.81E-14 & 0.2955 & 
1.22 &8360 &--- \\

IC 2003        & 03 56 22.0 & +33 52 28 & 4.97E-12 & 8.87E-14 & 3.40E-12 & 2.66E-13 & -1.26E-13 & 0.4451 & 
1.46 &1170--1950 &$>$1.8 \\
& & &          & 4.63E-14 &          & 1.39E-13 & -6.39E-14 &        & & & \\
 
NGC 1535       & 04 14 15.9 & -12 44 21 & 7.99E-12 & 1.06E-13 & 5.97E-12 & 3.17E-13 & -1.40E-13 & 0.5714 & 
1.34 &4550--5140 &$>$1.9 \\
& & &          & 6.35E-14 &          & 1.91E-13 & -6.00E-15 &        & & & \\

IC 418         & 05 27 28.2 & -12 41 50 & 1.59E-11 & 9.57E-14 & 1.42E-11 & 2.87E-13 & 1.61E-13 & 0.3658 & 
1.12 &12300--12600 &$>$3.4 \\ 
& & &          & 4.06E-14 &          & 1.22E-13 &  2.70E-14 &       &  & & \\

NGC 2022       & 05 42 06.2 & +09 05 10 & 3.56E-12 & 6.09E-14 & 2.40E-12 & 1.83E-13 & -5.99E-14 & 0.6721 & 
1.48 &660--1400 &$>$2.1  \\

IC 2149        & 05 56 24.0 & +46 06 15 & 2.96E-12 & 3.69E-14 & 2.26E-12 & 1.11E-13 & 2.12E-14 & 0.3213 & 
1.31 &5400--5960 &$>$1.9 \\ 
& & &          & 2.80E-14 &          & 8.39E-14 & -5.52E-15 &      & & & \\  
 
IC 2165        & 06 21 42.6 & -12 59 10 & 1.04E-11 & 2.05E-13 & 5.80E-12 & 6.14E-13 & -9.74E-14 & 0.4423 & 1.79 & out of range & \\
& & &          & 4.73E-14 &          & 1.42E-13 & -6.95E-14 &        &  & & \\

NGC 2346       & 07 09 22.5 & -00 48 24 & 7.01E-13 & 3.29E-14 & 4.95E-13 & 9.87E-14 & 2.66E-14 & 0.4853 & 
1.42 &2360 &--- \\ 
& & &          & 3.42E-14 &          & 1.02E-13 & 1.67E-14 &        &  & & \\

%%  NGC 2440       & 07 41 55.4 & -18 12 33 & 2.24E-11 & 1.48E-13 & 1.57E-11 & 4.45E-13 & 5.94E-13 & 0.5399 & 
{\bf NGC 2440}       & 07 41 55.4 & -18 12 33 & 2.24E-11 & 1.47E-13 & 1.57E-11 & 4.41E-13 & 5.94E-13 & 0.5399 & 
1.427 &2460 & {\bf 4.4$\pm$1.2} \\

PN G264.4-12.7 & 07 47 20.4 & -51 15 05 & 7.94E-13 & 1.92E-14 & 6.15E-13 & 5.76E-14 & 3.01E-14 & 0.2851 & 
1.29 &6000--7080 &$>$0.43 \\ 

IC 2448        & 09 07 06.6 & -69 56 29 & 8.06E-12 & 1.29E-13 & 5.00E-12 & 3.87E-13 & -1.06E-13 & 0.3831 & 
1.61 & out of range & \\
& & &          & 3.10E-14 &          & 9.30E-14 & 5.33E-14 &         & & & \\

NGC 2867       & 09 21 25.5 & -58 18 35 & 2.42E-11 & 1.26E-13 & 1.87E-11 & 3.77E-13 & 9.22E-14 & 0.4567 & 
1.29 &5900--6130 &$>$5.9 \\
& & &          & 9.29E-14 &          & 2.79E-13 &  4.31E-14 &       & & & \\

NGC 3132       & 10 07 01.8 & -40 26 11 & 1.71E-12 & 5.43E-13 & 1.08E-12 & 1.63E-12 & -1.33E-13 & 0.6262 & 
1.58 & out of range & \\
& & &          & 2.21E-13 &          & 6.63E-13 & -1.24E-13 &        &  & & \\

NGC 3211       & 10 17 50.4 & -62 40 12 & 7.94E-12 & 7.74E-14 & 6.04E-12 & 2.32E-13 & 2.24E-14 & 0.4637 & 
1.31 &5250--5680 &$>$2.8 \\ 

NGC 3242       & 10 24 46.1 & -18 38 33 & 4.71E-11 & 2.51E-13 & 3.15E-11 & 5.31E-13 & 5.13E-14 & 0.5662 & 
1.50 &370--530 &$>$14.5 \\
& & &          & 1.00E-13 &          & 3.00E-13 & 1.28E-13 &        & & & \\

NGC 3918       & 11 50 17.2 & -57 10 53 & 6.94E-11 & 2.16E-13 & 5.48E-11 & 6.48E-13 & 1.64E-13 & 0.4605 & 
1.27 &6810--6950 &$>$9.9 \\
& & &          & 7.69E-14 &          & 2.31E-13 & 1.51E-13 &        &  & & \\

NGC 4361       & 12 24 30.7 & -18 47 06 & 1.62E-12 & 6.75E-14 & 1.07E-12 & 2.02E-13 & -6.51E-14 & 0.6478 & 1.51 & out of range &\\

IC 3568        & 12 33 06.9 & +82 33 49 & 5.14E-12 & 2.83E-13 & 2.96E-12 & 8.49E-13 & 5.27E-14 & 0.3498 & 
1.74 & out of range & \\
& & &          & 3.46E-14 &          & 1.04E-13 & 4.53E-14 &        & & & \\

NGC 5189       & 13 33 31.7 & -65 58 23 & 3.77E-13 & 4.60E-14 & 3.00E-13 & 1.38E-13 & -6.37E-14 & 0.6241 & 
1.26 &7120 &--- \\

NGC 5315       & 13 53 55.2 & -66 30 52 & 1.08E-12 & 3.64E-14 & 1.90E-12 & 1.09E-13 & 1.06E-13 & 0.5320 & 
0.57 &58500 &--- \\

PN G342.1+27.5 & 15 22 19.3 & -23 27 46 & 1.06E-11 & 7.06E-14 & 7.25E-12 & 2.12E-13 & -4.57E-14 & 0.4228 & 
1.46 &1170--1470 &$>$6.9 \\

IC 4593        & 16 11 44.5 & +12 04 17 & 8.32E-13 & 4.29E-14 & 4.44E-13 & 1.29E-13 & -1.08E-14 & 0.3637 & 
1.87 & out of range & \\
& & &          & 4.07E-14 &          & 1.22E-13 & -9.61E-15 &        & & & \\ 

NGC 6210       & 16 44 29.5 & +23 48 00 & 6.74E-12 & 5.65E-14 & 5.28E-12 & 1.69E-13 & 1.53E-13 & 0.4977 & 
1.28 &6450--6820 &$>$3.1 \\ 
& & &          & 3.50E-14 &          & 1.05E-13 & 8.69E-14 &        &  & & \\

IC 4634        & 17 01 33.7 & -21 49 31 & 8.28E-13 & 4.42E-14 & 6.57E-13 & 1.33E-13 & -3.87E-14 & 0.4759 & 
1.26 &7010 &--- \\

NGC 6302       & 17 13 44.4 & -37 06 11 & 4.60E-13 & 3.24E-14 & 3.69E-13 & 9.72E-14 & 1.64E-14 & 0.4363 & 
1.25 &7450 &--- \\
& & &          & 3.57E-14 &          & 1.07E-13 & 8.36E-15 &        & & & \\

PN G345.2-08.8 & 17 45 35.3 & -46 05 23 & 1.34E-12 & 4.07E-14 & 1.06E-12 & 1.22E-13 & -7.79E-17 & 0.3101 & 
1.26 &6880--8240 &$>$0.08 \\
& & &          & 5.09E-14 &          & 1.53E-13 & -1.31E-14 &        &  & & \\

NGC 6543       & 17 58 33.4 & +66 37 60 & 9.25E-12 & 1.21E-13 & 7.38E-12 & 3.64E-13 & -6.68E-14 & 0.4973 & 
1.25 &7250--7830 &$>$1.5 \\
& & &          & 4.81E-14 &          & 1.44E-13 & -1.97E-14 &        & & & \\

NGC 6572       & 18 12 06.4 & +06 51 13 & 1.37E-11 & 1.07E-13 & 1.47E-11 & 3.20E-13 & -1.46E-13 & 0.3302 & 
0.93 &21900--22200 &$>$1.7 \\
& & &          & 2.23E-14 &          & 6.68E-14 & 6.38E-14 &        & & & \\

NGC 6644       & 18 32 34.7 & -25 07 44 & 1.13E-11 & 2.60E-13 & 8.13E-12 & 7.80E-13 & -4.17E-13 & 0.2668 & 
1.39 &3060--4080 &$>$0.82 \\
& & &          & 4.64E-14 &          & 1.39E-13 & -5.36E-14 &        &       & & \\

IC 1297        & 19 17 22.8 & -39 36 46 & 4.43E-12 & 1.12E-13 & 3.47E-12 & 3.35E-13 & -2.14E-13 & 0.4832 & 
1.28 &6450--7580 &$>$0.33 \\

NGC 6818       & 19 43 58.3 & -14 09 09 & 1.09E-11 & 2.81E-13 & 6.29E-12 & 8.42E-13 & -6.23E-13 & 0.5499 & 
1.73 & out of range & \\
& & &          & 4.04E-14 &          & 1.21E-13 & -3.31E-15 &        & & & \\

NGC 6826       & 19 44 48.2 & +50 31 30 & 1.15E-11 & 1.34E-13 & 8.21E-12 & 4.05E-13 & -3.86E-14 & 0.5048 & 
1.40 &2760--3280 &$>$2.8 \\
& & &          & 9.47E-14 &          & 2.84E-13 & 4.56E-14 &        & & & \\

NGC 6853       & 19 59 36.2 & +22 43 01 & 1.69E-12 & 5.09E-14 & 1.19E-12 & 1.53E-13 & -6.70E-14 & 0.6093 & 
1.42 &2250--3580 &$>$0.46 \\

NGC 6891       & 20 15 08.9 & +12 42 17 & 2.40E-12 & 3.81E-14 & 1.73E-12 & 1.14E-13 & 1.21E-14 & 0.3325 & 
1.39 &3140--3840 &$>$1.7 \\
& & &          & 4.02E-14 &          & 1.21E-13 & -3.20E-14 &       &  & & \\

NGC 6905       & 20 22 22.9 & +20 06 16 & 2.07E-12 & 6.85E-14 & 1.38E-12 & 2.06E-13 & -1.46E-13 & 0.9453 & 
1.50 &250 &--- \\

NGC 7009       & 21 04 10.8 & -11 21 49 & 1.49E-11 & 7.09E-14 & 1.22E-11 & 2.13E-13 & 1.92E-13 & 0.4601 & 
1.22 &8360--8570 &$>$5.6 \\
& & &          & 9.99E-14 &          & 3.00E-13 & 2.34E-13 &        &  & & \\

NGC 7027       & 21 07 01.6 & +42 14 10 & 3.64E-12 & 1.39E-13 & 4.88E-12 & 4.16E-13 & 2.07E-13 & 0.3556 & 
0.75 &36200 &--- \\
& & &          & 5.90E-14 &          & 1.77E-13 & 1.36E-13 &        & & & \\

PN G086.5-08.8 & 21 33 08.0 & +39 39 02 & 2.38E-12 & 4.70E-14 & 1.51E-12 & 1.41E-13 & -4.04E-14 & 0.6668 & 
1.58 & out of range & \\
& & &          & 5.35E-14 &          & 1.61E-13 & -3.49E-14 &        & & & \\
%% AKA HU2-1

IC 5217        & 22 23 55.8 & +50 57 60 & 8.79E-13 & 5.37E-14 & 6.66E-13 & 1.61E-13 & 1.01E-14 & 0.3017 & 
1.32 &5090 &--- \\ 

NGC 7662       & 23 25 54.0 & +42 32 06 & 4.55E-11 & 2.15E-13 & 3.53E-11 & 6.44E-13 & -2.49E-14 & 0.5772 & 
1.29 &6060--6270 &$>$6.5 \\
& & &          & 8.78E-14 &          & 2.63E-13 & 4.84E-14 &        &  & & \\

\enddata
\end{deluxetable}

\section{Rate Equations for the Level Populations}

In order to interpret the observational data,
we solve the \ion{C}{3} statistical equilibrium equations for the 
populations of the 4 lowest-lying energy levels, numbered 1--4.
These are $2s^2~^1S_0$,       
$2s2p~^3P_0$, $2s2p~^3P_1$, and $2s2p~^3P_2$.
For this, the effective collision strengths
($\Upsilon$ values) 
and transition
probabilities (A-values) listed in Table~2 are used.
Note the important distinction for A$_{21}$ between 
$^{12}$C and $^{13}$C.
This is the sole quantity that is different and
is the crucial factor in providing different results
when solving the rate equations for the level populations.
The effective collision strengths for transitions between
the $^3P_J$ levels are available for 
the electron temperature ($T_e$)~= 5000, 10000, 15000,
and 20000~K (Keenan et~al.\ 1992).
We use these same temperatures in Table~2, adding the recent
data for $^1S_0$--$^3P_J$ (Mitnik et~al.\ 2003 and 
Don Griffin, private communication). 
Then, following standard procedure, their value is partitioned
among the $^3P_J$ levels according to the statistical weights.

        The results of the solution of the rate equations for
$^{12}$C and for $^{13}$C in terms of the pertinent volume emissivity
($j$ value) ratios are given in Figure~2.
In this Figure, we plot $j_{1907}$/$j_{1909}$ for both pure
$^{12}$C and pure $^{13}$C in addition to
$j_{1910}$/$j_{1909}$ for pure $^{13}$C.
These ratios are shown vs.\ the log~($N_e$~cm$^{-3}$) for two values of $T_e$,
10000 and 15000~K.
This is a typical range for $T_e$ in PNs and demonstrates 
that there is minimal dependence on $T_e$.

        Besides the graphical representation shown in Figure~2,
it is useful to solve the 4-level rate equations explicitly in
the low-$N_e$ limit in terms of algebraic expressions for the atomic data. 
As will be seen, this provides superior insight when we compare the
observed line ratios with theoretical predictions.
The expression for the familiar $N_e$-diagnostic line ratio may be written,
\begin{equation} 
{I(1907) \over{I(1909)}} = {N_4~ A_{41}~ \nu_{41} \over{N_3~ A_{31}~ 
\nu_{31}}}~ .
\end{equation}
Here we need to obtain $N_4/N_3$.
How this was done for the general 3-level atom rate equations
may be seen from equations 2 and 3 in Rubin (1986).
Because we are writing the equations for just the 
low-$N_e$ limit,
when $A_{21}$ $\neq$ 0,
all the terms arising from levels above ground that contain the square of the density 
are dropped
(e.g., collisional routes out are negligible). 
Also terms with very low $A_{ij}$ may be neglected.
Then, in the low-$N_e$ limit when $A_{21}$ $\neq$ 0:
\begin{equation} 
{I(1907) \over{I(1909)}} = 1.001~{C_{14}~ A_{41} \over{C_{13}~ A_{41} + C_{14}~ 
A_{43}}}~ ,
\end{equation}
\noindent
where
$C_{ij}$~= $\Upsilon$$_{ij}$~$exp(-\chi_{ij}/kT_e)$.
The $\chi$s are the energy level differences and $k$ is
the Boltzmann constant. 
Because $A_{43} << A_{41}$, then
\begin{equation} 
{I(1907) \over{I(1909)}} = 1.001~{C_{14} \over{C_{13}}}~ .
\end{equation}

\begin{deluxetable}{cccccc}
\tabletypesize{\scriptsize}
\tablecaption{Effective Collision Strengths and Transition Probabilities 
for 4-Level \ion{C}{3} \label{tbl-2}}
\tablewidth{0pt}
\tablehead{
\colhead{} & \colhead{} & \colhead{  }{~~$T_e$} & \colhead{} & \colhead{} 
& \colhead{} \\
\colhead{i-j} & \colhead{5000} & \colhead{10000} & \colhead{15000} &
\colhead{20000} & \colhead{A$_{ji}$ s$^{-1}$}
}

\startdata                                          
1-2\tablenotemark{a} & 0.13 & 0.11222 & 0.10711 & 0.10556 & 0 (for $^{12}$C) \\
          &      &     &         &         & 9.04E-4 (for $^{13}$C)\tablenotemark{c}\\

1-3\tablenotemark{a} & 0.39 & 0.33667 & 0.32133 & 0.31667 & 103\tablenotemark{d,e} \\

1-4\tablenotemark{a} & 0.65 & 0.56111 & 0.53556 & 0.52778 & 5.14E-3\tablenotemark{c} \\

2-3\tablenotemark{b} & 0.800 & 0.950 & 1.01 & 1.03 & 2.39E-7\tablenotemark{f} \\

2-4\tablenotemark{b} & 0.590 & 0.695 & 0.797 & 0.855 & 1.4E-13\tablenotemark{c} \\

3-4\tablenotemark{b} & 2.23 & 2.75 & 3.05 & 3.23 & 2.39E-6\tablenotemark{c} \\

\enddata
\tablenotetext{a}{Don Griffin (private communication) using Mitnik et al. 2003
    partitioning among levels according to statistical weights}

\tablenotetext{b}{Keenan et al. 1992}

\tablenotetext{c}{Brage et al. 1998}

\tablenotetext{d}{Doerfert et al. 1997}

\tablenotetext{e}{J\"onsson \& Froese Fischer 1998}

\tablenotetext{f}{Fleming et al. 1996}

\end{deluxetable}

%%      ** Figure 2 about here **

\begin{figure}
\plotone{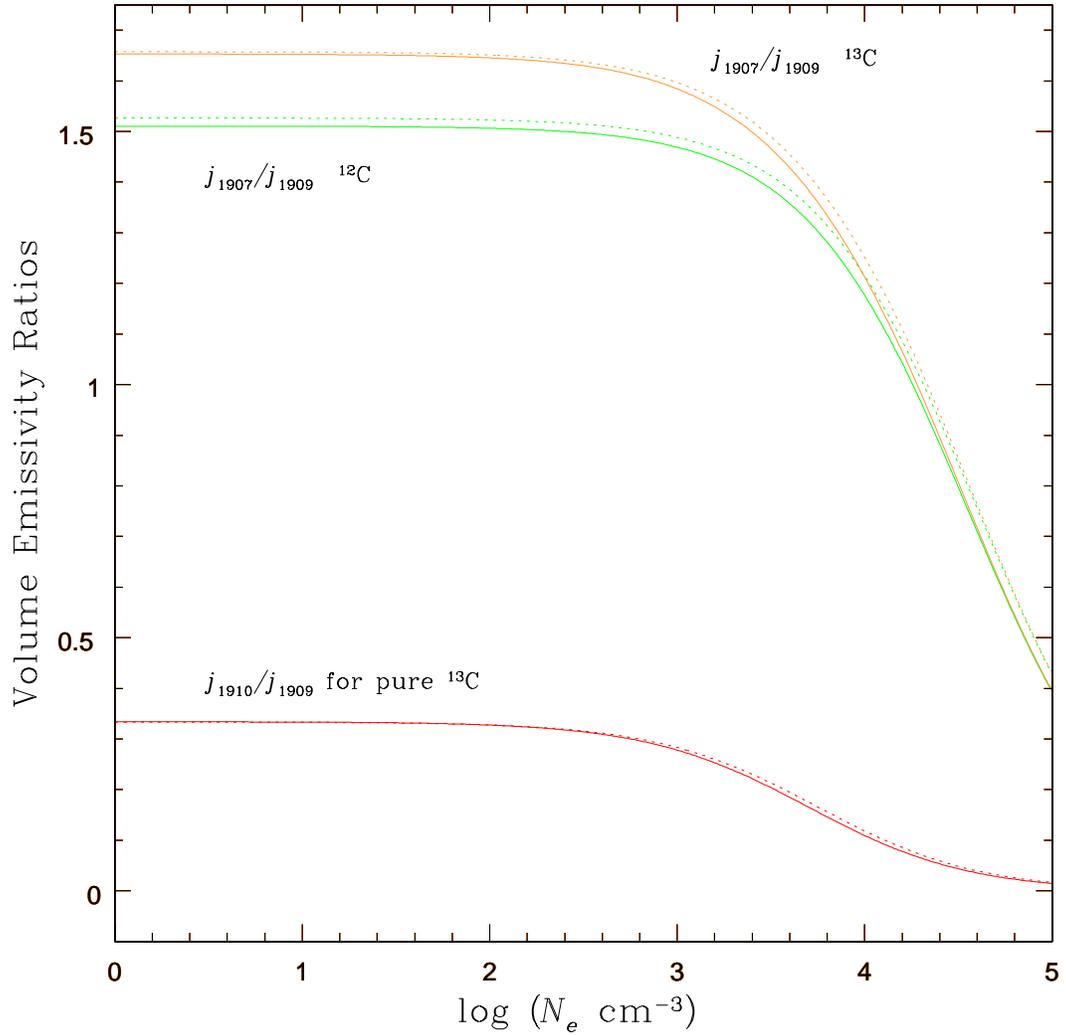}
\caption{The ratio of volume emissivities vs.\ log~($N_e$~cm$^{-3}$), 
from top to bottom:
$j_{1907}$/$j_{1909}$ for pure $^{13}$C and $^{12}$C  respectively,
$j_{1910}$/$j_{1909}$ for pure $^{13}$C. 
Note that the $j_{1909}$ values {\it differ} for the $^{12}$C and $^{13}$C 
curves.
The solid and dashed lines are for $T_e$~= 10000 and 15000~K.
\label{fig2}}
\end{figure}

\noindent
When we assume that all $exp(-\chi_{1j}/kT_e)$ are approximately the same, then
\begin{equation} 
{I(1907) \over{I(1909)}} \sim {\Upsilon_{14} \over{\Upsilon_{13}}} = {5 \over{3}}~ ,
\end{equation}
following the ratio of the statistical weights.
This is the equation governing the $^{13}$C  ratio.

        In the low-$N_e$ limit when $A_{21}$ = 0,
the collisional terms arising from level 2 are retained.
Thus, in the low-$N_e$ limit when $A_{21}$ = 0:
\begin{equation}
\frac{I(1907)}{I(1909)}  = 1.001~ \frac{C_{12}~\Upsilon_{24} + C_{14}~(\Upsilon_{23} + 
\Upsilon_{24} + \Upsilon_{12})}{C_{12}~\Upsilon_{23} + C_{13}~(\Upsilon_{23} + \Upsilon_{24}
+ \Upsilon_{12})}~.
\end{equation}
The last $\Upsilon_{12}$ term arises from collisional deexcitation 
out of level 2 and the symmetry of $\Upsilon_{21}$ and $\Upsilon_{12}$. 
Again, when all $exp(-\chi_{1j}/kT_e)$ are nearly the same, then
\begin{equation} 
\frac{I(1907)}{I(1909)} \sim \frac{\Upsilon_{12}~\Upsilon_{24} + \Upsilon_{14}~
(\Upsilon_{23} + \Upsilon_{24} + \Upsilon_{12})}{\Upsilon_{12}~\Upsilon_{23} + 
\Upsilon_{13}~(\Upsilon_{23} + \Upsilon_{24} + \Upsilon_{12})}~ .
\end{equation}
This is the equation governing the $^{12}$C  ratio.
We note that the second terms (including the parenthesized values)
in the numerator and denominator
dominate, and if the first terms were not present, would give the 
same result 5/3 as the $^{13}$C  ratio.
It is interesting that the two solutions hinge on just $A_{21}$.
Yet it is true that the low-$N_e$ limits for both the 
$^{12}$C and  $^{13}$C \ion{C}{3} ion depend on the collision
strengths.

        Let us now consider the ratio $I(1910)/I(1909)$ in the 
low-$N_e$ limit with $A_{21}$ $\neq$ 0.
This is for the case of pure $^{13}$C.
The bottom curve in Figure~2 and the behavior in the low-$N_e$ that we are
about to work out \underbar{do not} include any contribution
from $^{12}$C to the main lines.
Starting with
\begin{equation} 
\frac{I(1910)}{I(1909)} = \frac{N_2~ A_{21}~ \nu_{21}}{N_3~ A_{31}~ \nu_{31}}~,
\end{equation}
we find,
\begin{equation} 
\frac{I(1910)}{I(1909)} = 0.9995~ \frac{C_{12}~ A_{41} - 
C_{14}~ A_{43}}{C_{13}~ A_{41} + C_{14}~ A_{43}}~ .
\end{equation}
\noindent
The comments and assumptions that lead from equation (1) to (2) 
apply here, except that $N_2/N_3$ is apropos now.
Again, when all $exp(-\chi_{1j}/kT_e)$ are nearly the same, then
\begin{equation} 
\frac{I(1910)}{I(1909)} \sim \frac{\Upsilon_{12}~A_{41} - 
\Upsilon_{14}~ A_{43}}{\Upsilon_{13}~A_{41} + \Upsilon_{14}~A_{43}}~ .
\end{equation}
The second terms in the numerator and denominator
are negligible compared with the first terms;
by dropping them, one obtains
\begin{equation} 
\frac{I(1910)}{I(1909)} \sim \frac{\Upsilon_{12}}{\Upsilon_{13}} = 
\frac{1}{3}~ ,
\end{equation}
following the ratio of the statistical weights.

        The equations derived in this section agree extremely well
with the low-$N_e$ limits shown in Fig.~2.
The small variations with $T_e$ are also  
in agreement between the analytical expressions and Fig.~2.

\section{Interpretation of the Observed Line Fluxes}

        It is important to realize that the ``main" lines 1907 and 1909 are
produced by {\it both} $^{12}$C and $^{13}$C.
Following Brage et~al.\ (1998), we may write an equation for the
intrinsic line ratio $I(1910)/I(1909)$~= $I_{21}$/$I_{31}$  as
\begin{equation} 
\frac{I_{21}}{I_{31}} = \frac{\nu_{21}}{\nu_{31}} 
\frac{n_2 A_{21}}{(r m_3 + n_3) A_{31}}~ ,
\end{equation}
\noindent
where $r$ is the $^{12}$C/$^{13}$C abundance ratio,
$n_2$, $n_3$, and $n_4$ are the fractional populations for levels 2, 3, and 4 
for $^{13}$C and 
$m_3$ and $m_4$ are the fractional populations for levels 3 and 4 
for $^{12}$C. 
That is the sum over $n_i$, $i$~= 1--4 is unity, as is the sum over $m_i$.
Values of $n_i$ and $m_i$ are determined from solving the respective rate equations.

        We also write an analogous equation for the ratio of the        
main lines $I(1907)/I(1909)$~= $I_{41}$/$I_{31}$  as
\begin{equation} 
\frac{I_{41}}{I_{31}} = \frac{\nu_{41}}{\nu_{31}} \frac{(r m_4 + n_4) 
A_{41}}{(r m_3 + n_3) A_{31}}~ .
\end{equation}
For each of the sources in Table~1, there are measured flux ratios for
F(1907)/F(1909).
In addition, an upper limit to F(1910)/F(1909) may be set by using 3$\sigma$ 
as the upper limit for the non-detected 1910~\AA\ lines.
(Because lines in this multiplet are so close together,
intrinsic flux ratios are equivalent to observed flux ratios.)

        Equations (11) and (12) are two relations that depend on 
$r$ and $N_e$
with only a weak dependence on $T_e$.
For the solution of these equations, we use $T_e$~= 10000~K.
An iterative solution converged rapidly with the results
for $N_e$~cm$^{-3}$ and $r$, the $^{12}$C/$^{13}$C abundance ratio,
entered in the last two columns of Table~1.
Just as rapidly, there is an indication when there is no possible solution.
For these cases, a dash is shown in the 
$^{12}$C/$^{13}$C-column with the $N_e$ 
value found assuming pure $^{12}$C.
The reason that a range in $N_e$ is shown for those PNs with a
lower limit for 
$^{12}$C/$^{13}$C
is that the higher number for $N_e$ 
corresponds to the solution of
equations (11) and (12)
for that limiting value of $r$.
This value of $N_e$ is in fact an upper limit to the
actual $N_e$, which may range down to the smaller number shown
for pure  $^{12}$C. 

\subsection{The Flux Ratio F(1907)/F(1909)}

  For some of the PNs, 
we found that the measured
1907/1909~\AA\ flux ratio was higher than the
theoretical low--$N_e$ 
limit for $^{12}$C
1.511, 1.526, and 1.537 
at $T_e$ 10000, 15000, and 20000~K, respectively.
For instance NGC 6818 
has F(1907)/F(1909)~= 1.73, 
significantly higher than the atomic data predict. 
Rubin et~al.\ (2001)
found a similar problem for other $N_e$-diagnostic line ratios.
In the most recent work,
\citet{rub03} found 
that the observed 
[\ion{Ne}{5}] 14.3/24.3~$\mu$m 
flux ratio 
was out of theoretical bounds on the low-$N_e$ limit
for 10 out of 20 PNs 
in the Infrared Space Observatory (ISO) archive, including NGC 6818.

        A mixture of $^{13}$C with $^{12}$C helps to close the
gap somewhat.
The theoretical low--$N_e$ limit for $^{13}$C is 1.653, 1.658, and 1.660
at $T_e$ 10000, 15000, and 20000~K, respectively.
However, even if half the carbon were $^{13}$C, 
observed F(1907)/F(1909) ratios as high as 1.73 would not be explained.
Indeed any ratio exceeding the 5/3 value
would not comport with present predictions.
The value of 5/3 is the result of 
the standard practice of partitioning 
the $^1S_0$--$^3P_J$ multiplet value among the fine-structure levels
according to the statistical weights (1, 3, 5).
Consequently, this 5/3 limit does not 
depend on whose set of collision
strengths is employed as long as 
they are partitioned in this way.
% This limit does not depend on whose set of collision
% strengths is employed as long as the standard practice of partitioning 
% the $^1S_0$--$^3P_J$ multiplet value among the fine-structure levels
% according to the statistical weights (1, 3, 5)
% has been done.
An observational program to measure very accurate 
F(1907)/F(1909) ratios would be most useful.
On the theoretical side, a detailed calculation for the effective 
collision strengths between these individual $LS$ levels 
would be extremely valuable.

\begin{figure}[top]
\plotone{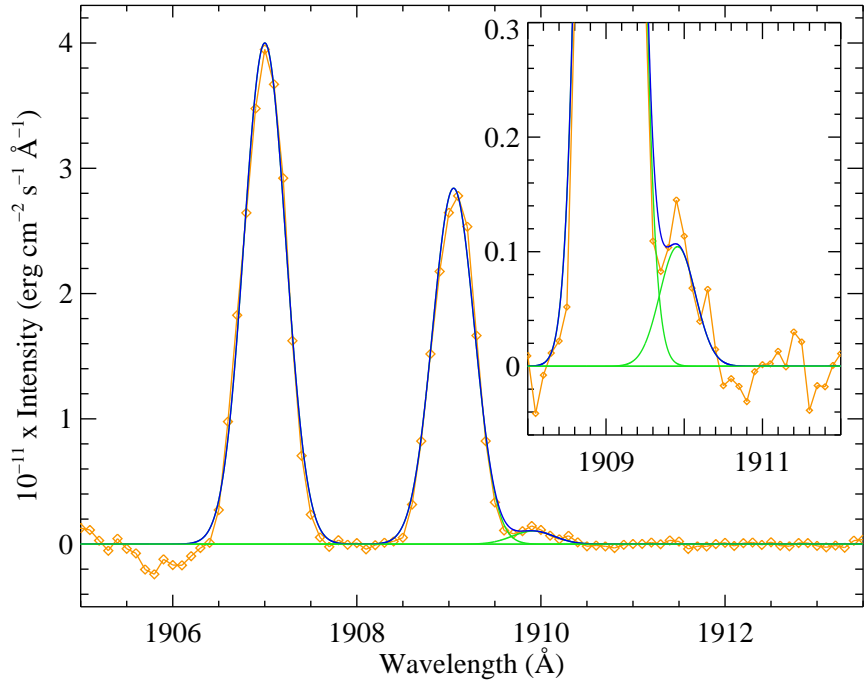}
\vskip-0.3truein
\caption{Spectrum in the vicinity of the 
\ion{C}{3} 
multiplet 
for NGC 2440 resulting from the coaddition
of IUE high dispersion, large aperture spectra
SWP~7263 and LWR~10741.
The continuum is very weak compared with the 1907 and 1909~\AA\
lines and has been fit with a quadratic baseline and removed.
The inset shows 
a magnified (in intensity) view  of the $^{13}$C 
1910~\AA\ line.
The 3-component Gaussian fit (see text)
and sum  are
shown, as well as the line connecting the data points (diamonds).
 \label{fig3}}
\end{figure}

\subsection{The Flux Ratio F(1910)/F(1909)-- Detection of the $^{13}$C line in NGC 2440}

  The IUE data and analysis presented in Table~1 show only one source, 
NGC~2440,
with a significant detection of the $^{13}$C line (Figure~3).
This is the coadded spectrum of two archival spectra SWP~7263 and LWR~10741.
The apertures for both spectra were centered at the same position.
The position angles of the aperture major axis
were 199.19 and 20.32$^{\rm o}$ respectively
(only $\sim$1$^{\rm o}$ mismatch)
thus providing nearly the same sky coverage.
Exposure times were 3600~s on 1979 November 29  
and 16200~s on 1981 May 29 respectively
and were observed under programs
NCBJH and NPDJH, PI J.\ Patrick Harrington.

        In the coadded spectrum the observed flux of the 1909.6~\AA\ line
is (5.94$\pm$1.47)$\times$10$^{-13}$
erg~cm$^{-2}$~s$^{-1}$,
which is a 4$\sigma$ detection.
The simultaneous solution of equations (11) and (12), 
assuming $T_e$~= 10000~K,
gives $N_e$~= 2460~cm$^{-3}$ and $^{12}$C/$^{13}$C~= 4.4.
With an assumed $T_e$ of 15000~K,
the solution becomes 
$N_e$~= 3130~cm$^{-3}$ and 
$^{12}$C/$^{13}$C~= 4.2.
We prefer the former solution because
according to figure~2 ($T_e$ vs.\ I.P.) in Bernard Salas et~al.\ (2002), 
it appears that the $T_e$ value
for the C$^{++}$ zone in NGC~2440 is closer to 10000 than 15000~K 
for the case C$^+$ $\rightarrow$ C$^{++}$ with ionization potential (I.P.)
24.4~eV.
Although not sensitive to the assumed $T_e$, the general trend we 
find here is that the higher the $T_e$, 
the higher the derived $N_e$ and the lower the derived 
$^{12}$C/$^{13}$C.
We estimate the uncertainty in the value of 
$^{12}$C/$^{13}$C as $\pm$1.2 from 
separate measurements and analysis of the two individual IUE spectra.

        In their study of NGC~2440, Bernard Salas et~al.\ 
used these same two IUE spectra.
According to their table~3,  
F(1907)~= 2.43$\times$10$^{-11}$ 
and F(1909)~= 1.77$\times$10$^{-11}$~erg~cm$^{-2}$~s$^{-1}$.
This leads to F(1907)/F(1909)~= 1.373,
while they have a slightly different ratio of 1.35
in their table~5 where they also list the derived 
$N_e$~= 4360~cm$^{-3}$ assuming $T_e$~= 15000~K.
We compare these to our entries in Table~1: 
F(1907)~= 2.24$\times$10$^{-11}$ 
and F(1909)~= 1.57$\times$10$^{-11}$~erg~cm$^{-2}$~s$^{-1}$.
This leads to F(1907)/F(1909)~= 1.427,
which will result in a lower derived $N_e$ irrespective of 
accounting for the effect of the presence of $^{13}$C.
It is not clear why their line fluxes are higher than ours.
It may be due to several factors.
For instance, they may have used the pre-NEWSIPS calibration.
They likely used a different technique than ours to form a weighted
average of the two IUE spectra.
Their fluxes may be taken directly from the Gaussian-fit areas
whereas we have used Gaussian fits for guidance but use the
option in ISAP [moment fit] and IRAF [splot e-option] to
measure the areas (except for measuring the 1910 line with the Gaussian).  
Generally we find that departures from
a Gaussian profile result in somewhat smaller line fluxes.
Additionally, the fitting of lines is not totally objective,
e.g, in the manner of accounting for the baseline, etc.

        Although the line was not detected in any of the other PNs, we 
were able to establish lower limits for the $^{12}$C/$^{13}$C ratio 
for many.
Even when coadding many spectra, the S/N was not high enough to
ascertain a useful value of 
$^{12}$C/$^{13}$C for most objects.  
HST studies similar to those of \citet{pal02} and \citet{cle97} could be used
to refine the limits presented here.

  At a late stage in our work on this project, we were surprised to
discover that a similar project had been undertaken by Miskey et~al.\
(2000).  They report a detection in NGC~2440, but appear to derive a much
higher $^{12}$C/$^{13}$C ratio of 39. They also report a detection in 
NGC~6302 with a ratio of 23.  
Unfortunately, since only an AAS abstract has
been published, their data and methods are not available for scrutiny.
We cannot report a detection of the 1910~\AA\ line in NGC~6302.
Two short wavelength spectra (SWP~10391 and SWP~8971) and one long wavelength 
spectrum (LWR 7722) are available for this nebula. 
Even the 1907, 1909 main lines are indistinguishable from the noise
in LWR~7722; coadding SWP~10391 and SWP~8971 produces a non-detection
of the 1910 line. 
SWP~10391 shows a small peak at the correct location, but at the same level 
as the noise;
thus a detection cannot be conclusively claimed.

  While ten SWP and eight LWP/LWR spectra were
available for NGC~3918, many of them showed saturation in the strong 
\ion{C}{3} lines, making them unusable for determining ratios.  
Similarly, all four available SWP spectra for NGC~3242 are saturated, and
the values from Table~1 come from only LWR~12705.  
Our limits are consistent with the findings 
of \citet{cle97} and \citet{pal02}.
We note that Palla et~al.\ 
used a 1$\sigma$ upper limit for 
F(1910), while we use 3$\sigma$.  
Using their 1$\sigma$ value 
and main line fluxes for NGC~3242 with  
our improved atomic data here,
we find that their 
$^{12}$C/$^{13}$C~$\geq$ 38 
becomes $\geq$ 44.4. 
However, when we use a 3$\sigma$ observed upper limit from their paper,
then more conservatively 
$^{12}$C/$^{13}$C~$\geq$ 14,
a value very similar to our limit in Table~1.

\section{Discussion and Conclusions}

        NGC~2440 has an unusually massive central star (CS), 0.72~M$_{\sun}$,
according to Wolff et~al.\ (2000),
and 0.932~M$_{\sun}$, according to
G\'orny et~al.\ (1997).
The progenitor star is believed to be more massive than 4~M$_{\sun}$, 
e.g.,  figure~3 in  Balser et~al.\ (2002) 
has $\sim$4.2--4.3~M$_{\sun}$ (see also the discussion in their 
\S2).
In their study of NGC~2440, Bernard Salas et~al.\ (2002) concluded,
based on a measured (C+N+O)/H ratio higher than that for NGC~7027,
that the CS must have been a more massive progenitor than was 
the star in NGC~7027, for which Bernard Salas et~al.\ (2001) found 
3-4~M$_{\sun}$.
Above 4~M$_{\sun}$, asymptotic giant branch (AGB) stars
develop deep convective envelopes with
very high base temperatures, leading to the operation of the CN-cycle; 
hence the name, hot bottom burning.
There, the CN-cycle operates at the base of the convective
envelope in a hydrogen-burning shell, where
$^{12}$C is converted into $^{13}$C and then into $^{14}$N.
This process would tend to drive 
$^{12}$C/$^{13}$C toward
the CN-cycle equilibrium ratio of $\sim$3.3 (Smith \& Lambert 1990;
Frost et~al.\ 1998).
When ratios close to the CN-cycle equilibrium value are measured 
in a star's atmosphere, it is indicative of the presence of nearly pure
CN-cycle material. 
Thus, the 
low $^{12}$C/$^{13}$C ratio 
of 4.4$\pm$1.2 that we find for NGC~2440 does support this picture.

        If the progenitor were less massive than 4~M$_{\sun}$, 
then the scenario changes.
Both the cool bottom processing (CBP) and standard models
predict 
$^{12}$C/$^{13}$C~
$\simeq$~20--30 for stars between 3--4~M$_{\sun}$ 
(Boothroyd \& Sackmann 1999; Renzini \& Voli 1981).
Then our derived 
$^{12}$C/$^{13}$C for NGC~2440 would not appear to
support either the standard or CBP  theories. 
While CBP does produce ratios as low as 4.4, it is only expected to occur in
progenitor stars of $\leq$2~M$_{\sun}$
(Wasserburg et~al.\ 1995). 
Two (unlikely) possibilities 
might allow agreement.
First, were the mass estimates for the CS progenitor too large
by a factor of $\sim$2,  
our $^{12}$C/$^{13}$C~= 4.4 for NGC~2440 would support the CBP theory. 
Second, perhaps CBP occurs in stars of 3--4~M$_{\sun}$.  
This is in fact contrary to the statement by Wasserburg et~al.\ (1995)
that no CBP is expected above $\sim$2.3~M$_{\sun}$. 
However a reconsideration of
model parameters might be useful to determine whether this could occur.

        The small $^{12}$C/$^{13}$C we find for NGC~2440 is
in the range derived by Balser et~al.\ (2002) in their study 
of molecular clouds associated with PNs.
As mentioned in the Introduction, from observations of
$^{12}$CO and $^{13}$CO millimeter lines, 
they found values of $^{12}$C/$^{13}$C from 2.2--31 in 9 PNs.
The extreme low value 2.2$\pm$0.03 was for the bipolar PN M1--16,
which is not on our list.
        We note that our small ratio for NGC~2440
is consistent with measurements of J-type carbon stars. 
Abia \& Isern (1997) found $<^{12}$C/$^{13}$C$>$~= 6$\pm$3
while 
Ohnaka \& Tsuji (1999) in their analysis of 
26 J-type C stars found an even lower average of 4.7$\pm$2.8 
with several ratios as low as $\sim$1--2, well below the equilibrium 
value for the CN-cycle.

  The question remains as to why NGC~2440 is the only nebula among those
we measured which has a detectable 1909.6~\AA\ $^{13}$C hyperfine structure
line, as well as why it has such a low 
$^{12}$C/$^{13}$C.  
Using IUE archival data, we also set 
upper limits on F(1909.6) in 40 other PNs available
and lower limits for $^{12}$C/$^{13}$C in 23 of these.
IUE had limited dynamic range that precludes setting more stringent
limits for most objects, though coadding spectra allows better S/N.  
NGC~2440 is among the brightest PNs in terms of either of the main line fluxes;
only NGC~2867,  NGC~3242, NGC~3918, and NGC~7662 are higher (see Table~1).
By using the measured 3$\sigma$ as an upper limit to their F(1909.6),
we derive lower limits on $^{12}$C/$^{13}$C for these four nebulae of respectively
5.9, 14.5, 9.9, and 6.5~-- already higher than what we find for 2440.
NGC~2440 is carbon rich with C/O of 1.9 (Bernard Salas et~al.\ 2002)
and a high excitation Peimbert type I PN (Hyung \& Aller 1998),
that is, it is a N- and He-rich PN.
The central star is among the hottest ($\sim$200,000~K) 
and most massive known (see earlier discussion).
In fact, in their entire sample of $\sim$80 PNs,
G\'orny et~al.\ (1997) 
derived a CS mass of 0.932~M$_{\sun}$,
which was the second largest of all.
This nebula also has $T_e$ reaching as high as 15000~-- 20000~K as 
measured in several high ionization species (Bernard Salas et~al.\ 2002).
NGC~2440 has a relatively low $N_e$ of 
2460~cm$^{-3}$ determined here (see Table~1).
Both lower $N_e$ and lower $^{12}$C/$^{13}$C lead to a higher
F(1910)/F(1909).
It may be that Type~I PNs have lower $^{12}$C/$^{13}$C ratios.  
Furthermore, NGC~2440 is a bipolar PN.
In their study, Balser et~al.\ (2002) found those PNs that 
have the lowest values of $^{12}$C/$^{13}$C are classified as bipolar.
Further theoretical and observational work is needed to
determine the cause of the low ratios.
An HST spectroscopic study of the \ion{C}{3} multiplet in
NGC~2440 would be desirable.

        Several of the PNs
have the measured F(1907)/F(1909) outside the 
low-$N_e$ theoretical limit for $^{12}$C.
Uncertainties in the measured fluxes
may account for some of these differences.
A mixture of $^{13}$C with $^{12}$C helps narrow the
gap somewhat.
Nevertheless, some of observed 
F(1907)/F(1909) ratios still appear too high to 
conform to the presently predicted limits.
We have shown that both the F(1907)/F(1909) 
and the F(1910)/F(1909) ratios
in the low-$N_e$ limit 
are predominantly influenced
by using the standard partitioning among the collision strengths
for the multiplet $^1S_0$--$^3P_J$ 
according to the statistical weights.
A detailed calculation for the effective collision
strengths between these individual levels would be valuable.
The determination of reliable $N_e$ values by this method as 
well as the $^{12}$C/$^{13}$C abundance ratio depends on it.
As far as we know, it appears that there have been no previous 
attempts to account for the presence and effects of $^{13}$C
when using the well-known \ion{C}{3} $N_e$-diagnostic 
F(1907)/F(1909).

\acknowledgments

Support for this work from a NASA 
Astrophysics Data Program (ADP) grant 
NAG5-13030
is gratefully acknowledged.
We thank the NASA Undergraduate Research Program (NASA-USRP),
which made 2003 summer work for EEC at NASA/Ames possible.
Valuable contributions were made by David Ho and Alessandra Radicati.
Thanks to Janet Simpson and
M\'onica Rodr\'\i guez
for reading the paper and helpful remarks.
We thank Myron Smith and Randy Thompson for 
providing useful comments on various aspects of the IUE spectra.
RHR acknowledges support from the Long-Term Space Astrophysics (LTSA)
program and thanks Scott McNealy for providing a Sun workstation.

\end{document}